\documentclass[prl,twocolumn,showpacs,preprintnumbers,amsmath,amssymb,floatfix]{revtex4-1}

\usepackage{graphicx}              
\usepackage{dcolumn}               
\usepackage{longtable}             
\usepackage{bm}                    
\usepackage{afterpage}
\usepackage{color}

\def\ket#1{\vert #1 \rangle}
\def\beq{\begin{equation}}
  \def\eeq{\end{equation}}
\newcommand{\matelem} [3]{\langle #1 | #2 | #3 \rangle}
\newcommand{\braket} [2]{\langle #1 | #2 \rangle}
\newcommand{\DC}{\mathcal{D}}
\newcommand{\SC}{\mathcal{S}}
\newcommand{\WC}{\mathcal{W}}
\newcommand{\TC}{\mathcal{T}}
\newcommand{\MC}{Monte Carlo}
\newcommand{\qMC}{quantum \MC}

\newcommand{\HF}{Hartree-Fock}
\newcommand{\ET}{E_{\rm T}}
\begin{document}

\title{Semistochastic Projector Monte Carlo Method}

\author{F. R. Petruzielo$^{1}$}
\email{frp3@cornell.edu}
\author{A. A. Holmes$^{1}$}
\email{aah95@cornell.edu}
\author{Hitesh J. Changlani$^{1}$}
\email{hjc55@cornell.edu}
\author{M. P. Nightingale$^{2}$}
\email{nigh@pobox.com}
\author{C. J. Umrigar$^{1}$}
\email{CyrusUmrigar@cornell.edu}
\affiliation{
  $^1$Laboratory of Atomic and Solid State Physics, Cornell University, Ithaca, New York 14853, USA\\
  $^2$Department of Physics, East Hall, University of Rhode Island, Kingston, Rhode Island 02881, USA\\
}

\date{\today}
\begin{abstract}
  We introduce a {\it semistochastic} implementation of the power
  method to compute, for very large matrices, the dominant eigenvalue and
  expectation values involving the corresponding eigenvector.
  The method is semistochastic in that the matrix
  multiplication is partially implemented numerically exactly and
  partially stochastically with respect to expectation values only.  Compared to a
  fully stochastic method, the semistochastic approach significantly
  reduces the computational time required to obtain the eigenvalue to
  a specified statistical uncertainty.  This is demonstrated by the
  application of the semistochastic \qMC\ method to 
  systems with a sign problem: the fermion Hubbard model and the carbon dimer.
\end{abstract}
\maketitle
\it{Introduction.\ ---\ }\rm Consider the computation of the dominant
eigenvalue of an $N\times N$ matrix, with $N$ so large that the matrix cannot
be stored.  Transformation methods cannot be used in this case, but
one can still proceed with the power method, also known as the
projection method,
as long as one can compute and store the result of multiplication of an arbitrary vector by the matrix.
When, for sufficiently large N, this is no longer feasible,
\MC\ methods can be used to represent stochastically both the
vector and multiplication by the matrix.  This suffices to
implement the power method to compute the dominant eigenvalue and
averages involving its corresponding eigenvector.

In this Letter, we propose a hybrid method consisting of numerically
exact representation and multiplication in a small {\it deterministic}
subspace, complemented by {\it stochastic} treatment of the rest of
the space.
This semistochastic projection method combines the advantages of both approaches: it greatly
reduces the statistical uncertainty of averages relative to purely stochastic
projection while allowing $N$ to be large.  These advantages are
realized if one succeeds in choosing a deterministic subspace that
carries a substantial fraction of the total spectral weight of the
dominant eigenstate.

Semistochastic projection has numerous potential applications:
transfer matrix Monte Carlo \cite{Nightingale1988} for classical
statistical mechanical systems, quantum Monte Carlo (QMC)
\cite{BlankenbeclerSugarPhysRevD.27.1304,Trivedi,NigUmr-BOOK-99}, and
the calculation of subdominant eigenvalues \cite{PhysRevB.62.1089}.

In this work we apply the semistochastic method to compute the ground
state energy of quantum mechanical Hamiltonians represented in a
discrete basis. In this context, deterministic projection is known as
Full Configuration Interaction (FCI) to chemists and as Exact
Diagonalization to physicists, whereas stochastic projection is the
essence of various projector QMC
methods~\cite{BlankenbeclerSugarPhysRevD.27.1304,Trivedi}.  Hence,
semistochastic projection shall be referred to as SQMC.  The benefit
of SQMC over the corresponding QMC method is large in many systems of
interest since the \HF\ determinant, augmented by a small set of
additional determinants, indeed represents a significant fraction of the
total spectral weight of the ground state wave function.

The Hamiltonians for the systems considered here suffer from a sign
problem, {\it i.e.,} no sign changes of basis states can be found that
render all off-diagonal matrix elements non-positive (which allows all
the coefficients of the desired eigenstate to be non-negative).  Until
recently, projector QMC had been used most successfully
for systems that do not have a sign
problem~\cite{BlankenbeclerSugarPhysRevD.27.1304,Trivedi}, or with an
uncontrolled, variational fixed-node approximation \cite{DutchGroup}.
The recent breakthroughs of Alavi and coworkers with their FCIQMC
method \cite{Booth2009} and its {\it initiator} extension
\cite{Cleland2010}, have enabled the treatment with a controllable
bias of matrices with a sign problem.
Consequently, the stochastic method to which we compare SQMC is
essentially the same as the initiator FCIQMC ({\it i}-FCIQMC) method
of Alavi with some minor differences as explained below.

\it{Theory.\ ---\ }\rm We start from an $N \times N$ Hermitian matrix
$H$, with eigenvalues $E_0 < E_1 \le \dots \le E_{N-1}$.  In our case,
$H$ is a Hamiltonian represented in an orthonormal basis
$\left\{\ket{\phi_1},\ldots,\ket{\phi_N}\right\}$.  To obtain the
lowest eigenvalue $E_0$, and its eigenvector $\psi^{(0)}$ with
components $\psi^{(0)}_i\equiv \braket{\phi_i}{\psi^{(0)}}$, we first
invert, shift and scale the Hamiltonian matrix:
\begin{equation}
  P=\openone+\tau(\ET\openone-H),
\end{equation}
where $\ET$ is a running estimate of $E_0$.

If $\ET=E_0$, $P$ has unit eigenvalue.  If $\tau<2/(E_{N-1}-E_0)$,
then the unit eigenvalue is the dominant one.
With $E_0$ unknown, $\ET$ is adjusted to ensure that the power method iterates
remain reasonably constant in norm.
When multiplication by $P$ is performed deterministically, the fastest
convergence rate is obtained for $\tau=2/(E_{N-1}-E_1)$; semi- or fully-stochastic
multiplications require smaller values of $\tau$ to reduce
the statistical noise \footnote{For each system studied, a
near-optimal value of $\tau$ can be determined from inexpensive
calculations with a small number of walkers.}.

Let $\chi^{(0)}$ be an arbitrary initial vector satisfying
$\braket{\chi^{(0)}}{\psi^{(0)}}\neq 0$.  Then, repeated application
of $P$ to $\chi^{(0)}$ yields
\begin{equation}
  \label{eqn:project_psi0}
  \chi^{(t+1)}=P\chi^{(t)}=P^{t+1} \chi^{(0)}.
\end{equation}
According to the power method, $\chi^{(M)}\propto\psi^{(0)}$ for
sufficiently large $M$.  If coefficients $w_i^{(t)}$ are defined by
the expansion
\begin{equation}
  \ket{\chi^{(t)}} = \sum_{i=1}^{N} w_i^{(t)} \ket{\phi_i},
\end{equation}
the semistochastic representation of the weights $w_i^{(t)}$ and the
multiplication by $P$ in Eq.~(\ref{eqn:project_psi0}) are defined as
follows.

Let $\DC$ be the set of indices of vector components treated
deterministically, and let $\SC$ be the set of those treated
stochastically, where $\DC\cup \SC=\{1,\dots,N\}$,
$\DC\cap\SC=\emptyset$, and $|D|\ll N$.  Accordingly, $P$ is the sum
of a deterministic block $P^{\DC}$, and a stochastic complement
$P^{\SC}$,
\begin{equation}
  P = P^{\DC} + P^{\SC},
\end{equation}
where
\begin{equation}
  P^{\DC}_{ij} =
  \begin{cases}
    P_{ij}, & \text{if }i, j\in \DC,  \\
    0  & \text{otherwise.}\\
  \end{cases}
\end{equation}

If the deterministic space is the entire space, then there is no sign
problem or statistical noise.  Consequently, we can expect that using
a deterministic subspace that is not the entire space will reduce the
sign problem and statistical noise.

The coefficients of the basis functions are represented as a
population of walkers. The number of walkers on an occupied
$\ket{\phi_i}$ is
\begin{equation}
  n_i = \mbox{max}(1, \lfloor\left|w_i\right| \rceil),
\end{equation} where $\lfloor \cdot \rceil$ denotes
the nearest integer and each walker has signed weight $w_i/n_i$.

Next, we proceed to the multiplication by $P$ which evolves the
coefficients from time $t$ to time $t+1$.
\begin{itemize}
\item To account for the off-diagonal elements in
  $P^{\SC}$, 
  for each walker on $\ket{\phi_i}$, a move to $\ket{\phi_j} \neq
  \ket{\phi_i}$ is made with probability $T_{ji}$.  A single walker on
  $\ket{\phi_i}$ contributes
  \begin{equation}
    \begin{cases}
      0,& i,j\in \DC\\
      \frac{\displaystyle P_{ j i}}{\displaystyle
        T_{ji}}\frac{\displaystyle w_{i}^{(t)}}{\displaystyle
        n_{i}^{(t)}} &\rm{otherwise}

    \end{cases}
  \end{equation}
  to the signed walker weight on $\ket{\phi_j}$.  The choice of $T$
  determines the probability that particular off-diagonal moves are
  made.  In this work, the near-uniform choice of Booth, Thom and
  Alavi is used \cite{Booth2009}.  To control sign problems present in
  our examples, we use the initiator idea \cite{Cleland2010}, which we
  generalized in that we increase the initiator threshold with the
  number of steps taken since the last visit to the deterministic
  space \cite{footnoteInitiator}.
\end{itemize}
\begin{itemize}
\item To account for the diagonal elements in
  $P^{\SC}$, 
  the contribution to the total signed walker weight on
  $\ket{\phi_j}$, with $j\in \SC$, is
  \begin{align}
    P_{jj}w_{j}^{(t)}.
  \end{align}
\item Deterministic evolution is performed with $P^{\DC}$.  The
  contribution to the signed weight on $\ket{\phi_j}$, with $j\in\DC$,
  is
  \begin{equation}
    \sum_{i \in \DC} P^{\DC}_{ji} w_{i}^{(t)}.
  \end{equation}
  $P^{\DC}$ is stored and applied as a sparse matrix.
\item Finally, for each $\ket{\phi_j}$, all signed walker weight
  generated on $\ket{\phi_j}$ is summed, taking into account the sign
  of the contribution.  To avoid the large computational and memory
  cost of having small weights on a large number of basis states,
  basis states with weight less than some minimum cutoff, $w_{\rm
    min}$, are combined via an unbiased prescription
  \cite{UmrNigRun-JCP-93}.
\end{itemize}

After sufficiently many multiplications by $P$, contributions from
subdominant eigenvectors die out on average.  At this point, the
collection of averages begins.  The most commonly employed estimator
for the dominant eigenvalue is the {\it mixed estimator}
\begin{equation}
  E_{\rm{mix}} = \frac{ \matelem{\psi^{(0)}}{\hat{H}}{\psi_T}  }{\braket{\psi^{(0)}}{\psi_T}  },
\label{eqn:e_mix}
\end{equation}
where the trial state $\ket{\psi_T}$ satisfies
$\braket{\psi^{(0)}}{\psi_T}\neq 0$.

The trial state $\ket{\psi_T}$ is a linear combination of basis states
\footnote{In FCIQMC \cite{Booth2009,Cleland2010}, a single state, the
  \HF\ determinant, has been used as the trial state.},
\begin{equation}
  \ket{\psi_T} = \sum_{i \in \TC} d_{i} \ket{\phi_i}, \label{eqn:psit_linear_combination}
\end{equation}
where $\TC$ is the set of indices of those basis functions that
contribute to the trial state.  We require that $|\TC|\ll N$, but not
necessarily that $\TC \subset \DC$.

At any particular time $t$, the stochastic representation of the
dominant eigenvector is
\begin{equation}
  \ket{\psi^{(0)}}  \approx \ket{\chi^{(t)}} = \sum_{i \in \WC^{(t)}} w_{i}^{(t)} \ket{\phi_i},
\end{equation}
where $\WC^{(t)}$ is the set of indices of basis functions occupied by
walkers at time $t$.  The full representation of the dominant
eigenvector is obtained by averaging over Monte Carlo generations
\begin{equation}
  \ket{\psi^{(0)}} \approx
  \frac{1}{N_{\rm{gen}}}\sum_{t=1}^{N_{\rm{gen}}} \sum_{i\in \WC^{(t)}} w_{i}^{(t)} \ket{\phi_i},
\end{equation}
where $N_{\rm{gen}}$ is the number of times $P$ is applied
after equilibration.

For the trial state in Eq.~(\ref{eqn:psit_linear_combination}),
$E_{\rm{mix}}$ of Eq.~(\ref{eqn:e_mix}) is
\begin{equation}
  E_{\rm{mix}} = \frac{\sum_{t=1}^{N_{\rm{gen}}} \sum_{i\in \WC^{(t)}}  w_{i}^{(t)} \sum_{j \in \TC}  H_{ij} d_{j}}
  {\sum_{t=1}^{N_{\rm{gen}}} \sum_{i\in \WC^{(t)} \cap \TC}  w_{i}^{(t)}  d_{i}}.
\end{equation}
Since $E_{\rm{mix}}$ is a zero-variance and zero-bias estimator when
$\ket{\psi_T}$ is equal to the dominant eigenvector, improving the
quality of $\ket{\psi_T}$ reduces fluctuations and bias in the mixed
estimate of the dominant eigenvalue.  This reduction can be achieved
with almost no additional computational cost by storing nonzero
$\sum_{j \in \TC} H_{ij} d_{j}$ terms.

\it{Applications.\ ---\ }\rm The semistochastic method is now applied
to compute the ground state energy of
the carbon dimer and the simple-square $8 \times 8$ fermionic Hubbard model with
periodic boundaries.
In both cases, we represent $H$ in the basis of determinants formed
from the restricted Hartree-Fock orbitals.  For the Hubbard model these
orbitals are the momentum eigenstates.  For the carbon dimer these orbitals
are obtained by solving the restricted Hartree-Fock equations in
cc-pVTZ basis set \cite{DunningJr1989}.
The majority of the Hubbard calculations are performed for $U/t=4$,
where $U$ is the on-site Coulomb repulsion and $t$ is the nearest
neighbor hopping parameter.  This parametrization is considered to be in the
{\it intermediate coupling} regime (the noninteracting bandwidth being
$8t$), and has been used widely in the literature
\cite{Hubbard}.

The trial wave function space and the deterministic space are generated
with identical iterative schemes, but possibly different parameters.
At each iteration, first define a reference space as all states
obtained in the previous iteration.  Second, generate a space which
includes all determinants connected to the reference space by a single
application of the Hamiltonian.  Third, find the dominant eigenvector
in this space.  Fourth, truncate the space using a criterion based on
the magnitude of the coefficient of each state in the eigenvector.
This truncated space becomes the reference for the next iteration.
The reference for the first iteration is the Hartree-Fock state.

For various sizes of the deterministic space, we demonstrate the
improvements of SQMC over the purely stochastic method defined by a
deterministic space which includes only the Hartree-Fock determinant.
The purely stochastic method is almost the same as {\it
  i}-FCIQMC~\cite{Booth2009,Cleland2010}, aside from some details such
as the use of real walker weights versus the integer walker weights
used in FCIQMC and the use of a graduated initiator in
SQMC~\cite{footnoteInitiator}.  The most dramatic benefit of SQMC is
in the efficiency, which is defined to be proportional to the inverse
of the time required to obtain the ground state energy to a specified
level of uncertainty.

\begin{figure}[htp] \begin{center}
    \includegraphics[scale=0.70]{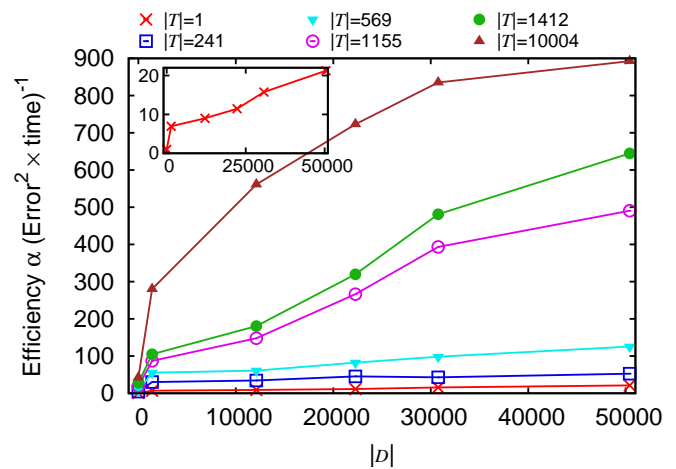}
    \caption{Relative efficiency of SQMC {\it vs.} dimension $|\DC|$ of the
      deterministic space for the simple-square $8 \times 8$ Hubbard
      model with periodic boundaries, $U/t=4$ and 10 electrons.
      Results are shown for trial wave functions of increasing size.
      The inset shows the
      $|\TC|=1$ curve on an expanded scale.  For this system, $N \approx 10^{12}$.
      }
    \label{fig:efficiency_8855}
  \end{center}
\end{figure}

To show the gain in efficiency of SQMC we computed the relative
efficiency, {\it i.e.}, the efficiency normalized by that of the
stochastic method ($|\DC|=1$), with $|\TC|=1$.
Fig.~\ref{fig:efficiency_8855} shows the relative efficiency of SQMC {\it
  vs.} the size of the deterministic space for the simple-square $8
\times 8$ Hubbard model with periodic boundaries, $U/t=4$ and 10
electrons.  The orders of magnitude increases in efficiency
demonstrate the benefits not only of SQMC but also of
improving the trial wave function.  The gain
of just using the largest deterministic space is a factor of
22, while the benefit of just using the largest trial wave function is
a factor of 42.  Both together yield a factor of about 900 as seen in the plot,
but the two are not always multiplicative.

\begin{figure}[htp] \begin{center}
    \includegraphics[scale=0.70]{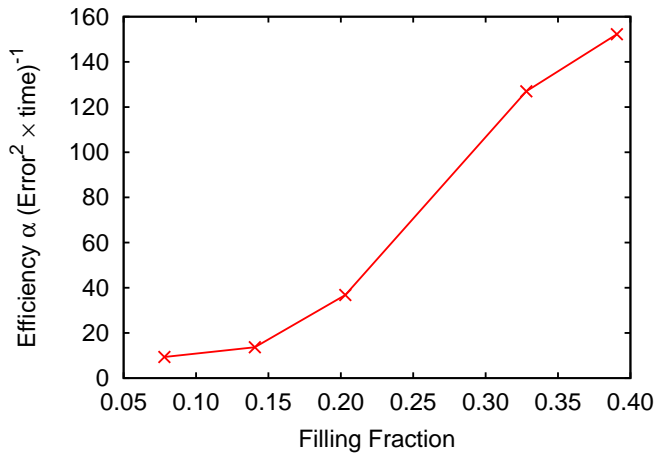}

    \caption{Relative efficiency of the SQMC {\it vs.} filling fraction
      for the simple-square $8 \times 8$ Hubbard model with $U/t=4$.
      In all cases, the trial wave function is the Hartree-Fock determinant.
      The deterministic space is constructed by applying the Hamiltonian once to the Hartree-Fock determinant. This yields
      spaces of sizes 1412, 4088, 7424, 14160, 16540.
      $N$ ranges from roughly $10^{12}$ to $10^{35}$.}
    \label{fig:efficiency_hubbard}
  \end{center}
\end{figure}

Fig.~\ref{fig:efficiency_hubbard} shows the efficiency gain of SQMC
{\it vs.} filling fraction for the simple-square $8 \times 8$ Hubbard
model with $U/t=4$.
The deterministic space, constructed by applying the Hamiltonian once to the Hartree-Fock determinant,
has a rather modest increase in size from 1412 to 16540 determinants,
whereas the size of the Hilbert space grows enormously from about $10^{12}$ to $10^{35}$.
Nevertheless, the efficiency gains increase with filling fraction.
Calculations beyond the scope of the present paper show that the initiator bias, at all fillings, decreases
with increasing $\DC$, but that it increases with filling fraction and $U$ in both the stochastic and the semistochastic methods.

\begin{figure}[htp] \begin{center}
    \includegraphics[scale=0.70]{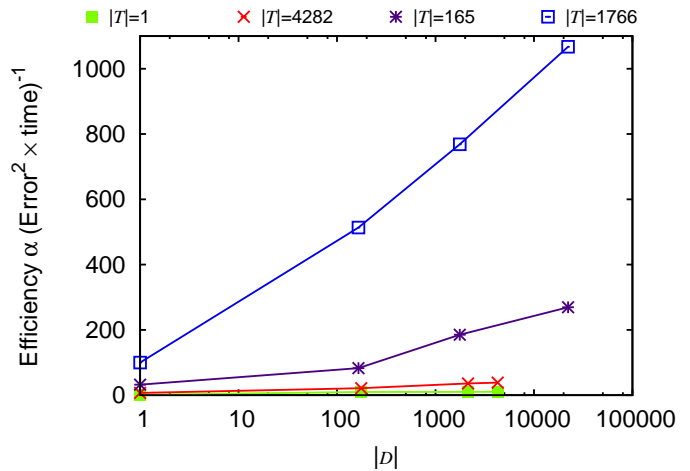}
     \caption{Relative efficiency of SQMC {\it vs.} dimension $|\DC|$ of the deterministic space
       for the carbon dimer with a cc-pVTZ basis.
       Results are shown for trial wave functions of increasing size.
       The top two curves are for $\DC$ and $\TC$ generated with two applications of our iterative scheme.
       The 165 and 1766 determinant wave functions with some quadruple excitations
       have much higher efficiency than the 4282 determinant wave function without any.
       For this system, $N \approx 10^{9}$.
       }
    \label{fig:efficiency_c2}
  \end{center}
\end{figure}

SQMC produces large efficiency gains for chemical systems as well.
Fig.~\ref{fig:efficiency_c2} shows the efficiency gain of SQMC {\it
  vs.} the size of the deterministic space for the carbon dimer with a
cc-pVTZ basis set \cite{DunningJr1989}.
The bottom two curves are for $\DC$ and $\TC$ generated with one
applications of our iterative scheme which generate single and double
excitations only.  The largest efficiency gain for these is about 40.
The top two curves are for $\DC$ and $\TC$ generated with two
applications of our iterative scheme and, hence, include several
chemically relevant quadruple excitations which are important for
correctly describing the ground state wave function.  The largest
efficiency gain now jumps to over 1000.

\begin{figure}[htp] \begin{center}
    \includegraphics[scale=0.50]{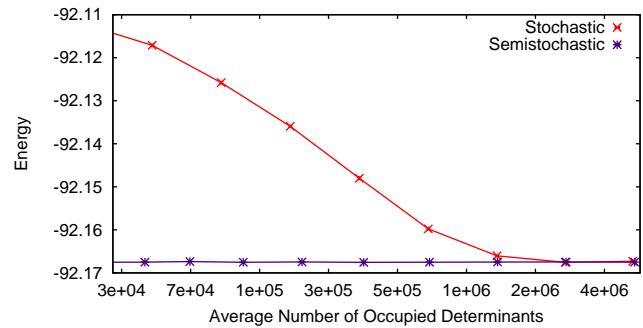}
    \caption{Energy of SQMC and the stochastic method {\it vs.}  the
      average number of occupied determinants for the simple-square $8
      \times 8$ Hubbard model with $U/t=1$ and 50 electrons.  The
      trial wave function for each of these calculations is the
      Hartree-Fock determinant.
      The deterministic space consists of the 16540 determinants connected to the Hartree-Fock determinant.
      For this system, $N\approx 10^{35}$.}
    \label{fig:bias_u_1_882525.eps}
  \end{center}
\end{figure}

Not only is SQMC much more efficient than the stochastic method, but
in some cases, also the initiator bias is significantly reduced.
Fig.~\ref{fig:bias_u_1_882525.eps} shows the biased estimates of the
energy as obtained by both the SQMC and stochastic method {\it vs.}
the average number of occupied determinants for the $8 \times 8$
Hubbard model with $U/t=1$ and 50 electrons.  SQMC has essentially no
bias.  A larger average number of occupied determinants corresponds to
using a larger walker population in the calculation.  The time
required for a calculation is proportional to the walker population.

\begin{figure}[htp] \begin{center}
    \includegraphics[scale=0.50]{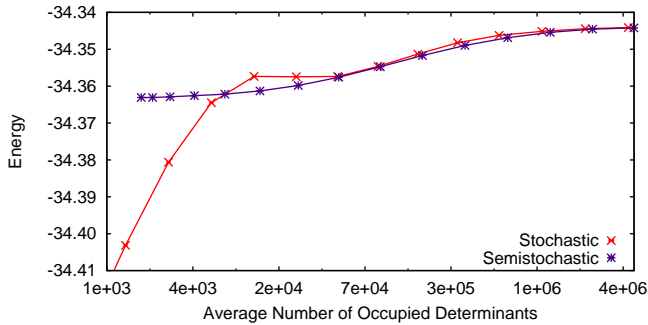}
    \caption{Energy of SQMC and the stochastic method {\it vs.}  the
      average number of occupied determinants for the simple-square
      $8 \times 8$ Hubbard model with $U/t=4$ and 10 electrons.
      The trial wave function for each of these calculations is the Hartree-Fock determinant.
      The deterministic space reference state for
      each SQMC calculation is the \HF\ determinant, yielding a
      deterministic space of 1412 determinants.
      For this system, $N\approx 10^{12}$.}
    \label{fig:bias_u_4_8855.eps}
  \end{center}
\end{figure}

The reduction in initiator bias is not always large.
Fig.~\ref{fig:bias_u_4_8855.eps} shows both the SQMC and stochastic
method energy {\it vs.} the average number of occupied determinants
for the $8 \times 8$ Hubbard model with $U/t=4$ and 10 electrons.
SQMC has a reduced initiator bias for a small, but not for a large
number of occupied determinants.  However, for this system and all
other systems studied SQMC has a smoother bias than the stochastic
method.

\it{Conclusion.\ ---\ }\rm The semistochastic power method, a hybrid
with deterministic and stochastic components, was introduced for
finding the dominant eigenvalue and sampling the corresponding
eigenvector of a matrix.  We showed that this novel, deterministic
component significantly reduces the
noise of the purely stochastic method without compromising its ability
to deal with matrices well beyond the size that can be handled
by purely deterministic methods.  In particular, matrices ranging in order
from $10^{9}$ to $10^{35}$ were successfully tackled. 
Besides being more efficient than a purely stochastic approach, the
semistochastic method has in some cases the additional benefit of a
much reduced initiator bias.
Also, the bias tends to be smoother and more amenable to removal by extrapolation.
We only presented applications to
systems with a sign problem, but the efficiency benefits of a
semistochastic implementation of the power method extend to systems
without a sign problem.

\it{Acknowledgments.\ ---\ }\rm We thank Garnet Chan and Ali Alavi for valuable discussions.
This work was supported in part by DOE-CMCSN DE-SC0006650 (AAH), NSF CHE-1112097 (HJC) and NSF DMR-0908653 (FRP).

\bibliographystyle{apsrev4-1}
\bibliography{semistoch}

\begin{thebibliography}{10}
\bibitem{footnoteInitiator}%
  \BibitemOpen
  \bibinfo {note} {In the initiator approach, a state must have a minimum
  absolute weight $i$ in order to contribute an off-diagonal move to a state
  that is not already occupied. This results in a non-variational bias that
  vanishes in the limit of infinite walker population. In our modification of
  this approach, $i=cm^p$, where $m$ is the number of moves the walker has made
  since its last visit to the deterministic space and $c$ and $p$ are
  constants, chosen to be 1 for the applications in this paper.}%
  \bibAnnoteFile{Stop}{footnoteInitiator}%
\end{thebibliography}
\end{document}